\documentclass[a4paper,11pt]{article}
\usepackage{jinstpub} 
\usepackage{lineno}
\usepackage{booktabs}
\usepackage{orcidlink}
\usepackage[misc]{ifsym}

\title{\boldmath Characterization of field cage and cathode for low radioactivity operation with the CYGNO experiment}

\author[1]{F.D. Amaro \orcidlink{0000-0001-7315-0550}}
\author[2,3]{R. Antonietti \orcidlink{0009-0009-2568-8247}}
\author[4,5]{E. Baracchini \orcidlink{0000-0003-4686-128X}}
\author[6]{L. Benussi \orcidlink{0000-0002-2363-8889}}
\author[6]{S. Bianco \orcidlink{0000-0002-8300-4124}}
\author[6,7,+,*]{A. Biondi \orcidlink{0009-0009-9199-7814}}
\author[6]{C. Capoccia \orcidlink{0009-0008-5919-3130}}
\author[6,8]{M. Caponero \orcidlink{0000-0002-5728-3123}}
\author[9]{L.G.M de Carvalho \orcidlink{0009-0003-5836-4771}}
\author[7,10]{G. Cavoto \orcidlink{0000-0003-2161-918X}}
\author[6]{I.A. Costa \orcidlink{0000-0002-3064-8305}}
\author[6]{A. Croce}
\author[4,5]{M. D'Astolfo \orcidlink{0009-0000-9817-6693}}
\author[10]{G. D'Imperio \orcidlink{0000-0002-2945-0983}}
\author[6]{E. Dan\`e \orcidlink{0000-0002-7220-9984}}
\author[6,*]{G. Dho \orcidlink{0000-0001-9454-9894}}
\author[10]{E. Di Marco \orcidlink{0000-0002-5920-2438}}
\author[1]{J.M.F. dos Santos \orcidlink{0000-0002-8841-6523}}
\author[4,5]{D. Fiorina \orcidlink{0000-0002-7104-257X}}
\author[10]{F. Iacoangeli \orcidlink{0000-0003-0808-585}}
\author[4,5]{Z. Islam \orcidlink{0000-0003-4611-839X}}
\author[11]{E. Kemp \orcidlink{0000-0001-5311-1300}}
\author[4,5]{H. P. Lima Jr \orcidlink{0000-0001-7398-3237}}
\author[6]{G. Maccarrone \orcidlink{0000-0002-7234-9522}}
\author[1]{R.D.P. Mano \orcidlink{0000-0003-2920-7067}}
\author[4,5]{D. J. G. Marques \orcidlink{0000-0002-0013-6341}}
\author[6]{G. Mazzitelli \orcidlink{0000-0003-2830-4359}}
\author[12]{A.G. McLean}
\author[2,3]{P. Meloni \orcidlink{0009-0001-7634-370X}}
\author[7,10]{A. Messina \orcidlink{0000-0003-1195-6780}}
\author[1]{C.M.B. Monteiro \orcidlink{0000-0002-1912-2804}}
\author[9]{R.A. Nobrega \orcidlink{0000-0001-5199-308X}}
\author[9]{I.F. Pains \orcidlink{0009-0004-0851-6308}}
\author[6]{E. Paoletti}
\author[6]{L. Passamonti}
\author[2,3]{F. Petrucci \orcidlink{0000-0002-5278-2206}}
\author[4,5]{S. Piacentini \orcidlink{0000-0002-1256-7149}}
\author[6]{D. Piccolo}
\author[6]{D. Pierluigi}
\author[10]{D. Pinci \orcidlink{0000-0002-7224-9708}}
\author[4,5]{A. Prajapati \orcidlink{0000-0002-4620-440X}}
\author[10]{F. Renga \orcidlink{0000-0001-8129-8504}}
\author[6]{F. Rosatelli}
\author[6]{A. Russo}
\author[6,13]{G. Saviano}
\author[1]{P.A.O.C. Silva \orcidlink{0000-0002-1957-2274}}
\author[12]{N.J. Spooner}
\author[6]{R. Tesauro \orcidlink{0009-0006-0722-5896}}
\author[6]{S. Tomassini \orcidlink{0000-0001-7290-2028}}
\author[4,5]{S. Torelli \orcidlink{0000-0003-3622-3524}}
\author[7,10]{D. Tozzi \orcidlink{0009-0001-9206-7354}}
\affiliation[1]{LIBPhys; Department of Physics; University of Coimbra; 3004-516 Coimbra; Portugal}
\affiliation[2]{Dipartimento di Matematica e Fisica; Universit\`a Roma TRE; 00146; Roma; Italy}
\affiliation[3]{Istituto Nazionale di Fisica Nucleare; Sezione di Roma Tre; 00146; Rome; Italy}
\affiliation[4]{Gran Sasso Science Institute; 67100; L'Aquila; Italy}
\affiliation[5]{Istituto Nazionale di Fisica Nucleare; Laboratori Nazionali del Gran Sasso; 67100; Assergi; Italy}
\affiliation[6]{Istituto Nazionale di Fisica Nucleare; Laboratori Nazionali di Frascati; 00044; Frascati; Italy}
\affiliation[7]{Dipartimento di Fisica; Universit\`a di Roma Sapienza; 00185; Roma; Italy}
\affiliation[8]{ENEA Centro Ricerche Frascati; 00044; Frascati; Italy}
\affiliation[9]{Universidade Federal de Juiz de Fora; Faculdade de Engenharia; 36036-900; Juiz de Fora; MG; Brasil}
\affiliation[10]{Istituto Nazionale di Fisica Nucleare; Sezione di Roma; 00185; Roma; Italy}
\affiliation[11]{Universidade Estadual de Campinas  - UNICAMP;  Campinas 13083-859; SP; Brazil}
\affiliation[12]{Department of Physics and Astronomy; University of Sheffield; Sheffield; S3 7RH; UK}
\affiliation[13]{Dipartimento di Ingegneria Chimica; Materiali e Ambiente; Sapienza Universit\`a di Roma; 00185; Roma; Italy}
\affiliation[+]{Now at Jagiellonian University, Doctoral School of Exact and Natural Sciences; 30-348; Cracow; Poland }
\affiliation[*]{Corresponding}

\emailAdd{alex.biondi@doctoral.uj.edu.pl}
\emailAdd{giorgio.dho@lnf.infn.it}

\abstract{
Dark matter, which is considered to account for approximately the 27\% of the Universe’s energy-mass content, remains an open issue in modern particle physics along with its composition. The CYGNO Experiment aims to exploit an innovative approach applied to the direct detection search of low energy nuclear recoils possibly induced by cold particle-like dark matter candidates. CYGNO employs a directional detector based on a Time Projection Chamber (TPC) filled with a He:CF$_{4}$ gas mixture and equipped with an optical readout. Currently, the CYGNO Collaboration is constructing the detector demonstrator, CYGNO-04, in Hall F at Laboratori Nazionali del Gran Sasso (LNGS). This 0.4 m$^3$ detector has the goal of proving the scalability of the technology and assessing the physics and radiopurity capabilities. Given the low radioactivity requirements, especially in internal components such as field cage and cathode, the reduction of material while keeping the correct electrical behavior is paramount. In this paper, we present the validation of several internal components, mainly focusing on the field cage material and support structure. The tests included geometrical asymmetries in the electric field response, collection efficiency as well as measurement of known physical quantities. A preferred configuration is found with a structure based on Nylon material which supports a PET or Kapton sheet with copper strips deposited on. }

\begin{document}
\maketitle
\flushbottom

\section{Introduction}
\label{sec:introduction}
Dark matter (DM) remains one of the most significant unresolved questions in modern physics. Its possible presence is suggested by a plethora of astrophysical, astronomical and cosmological observations which probed our Universe at different scales. In the current leading theories, DM is considered to constitute approximately 27\% of the Universe’s total mass-energy content and it interacts with ordinary matter only through gravity and possibly another unknown weak interaction \cite{bertone2005particle}. Despite extensive research, the composition of DM remains unknown.\\
One of the leading hypotheses is that it consists of Weakly Interacting Massive Particles (WIMPs) or WIMP-like particles. These hypothetical particles can naturally arise in several extensions of the Standard Model of particle physics and can be thermally produced in the early Universe to match the abundance estimated today. They are characterized by extremely weak interaction with Standard matter, by being neutral, non-baryonic and cold, i.e. non-relativistic at the decoupling phase with regular matter. 
As DM particles are expected to populate a halo around our Galaxy, a possible method to detect them is to directly observe the rare standard matter recoils induced by them. Direct detection experiments are exploring this possibility, mostly focusing on nuclear recoils (NRs) for kinematic reasons, in a theoretical justified range of DM masses from hundreds of MeV/c$^2$ up to TeV/c$^2$ scale \cite{zurek2014asymmetric,petraki2013review,hochberg2015model,xenon2023,crest2024,supercdms2024,lz2024,pandax2024,darkside2023,newsg2024}.
Moreover, the motion of the Solar System through the Galaxy with a speed similar to the average one of DM particles induces a peculiar dipolar structure to the angular distribution of the NRs. This feature is considered key for further improvements on the field, from the possibility of a positive claim of Galactic DM discovery, to search within the neutrino fog and DM astronomy \cite{mayet2016review,ohare2021fog,vahsen2021directional}.\\
The CYGNO experiment \cite{amaro2022cygno}, part of the broader CYGNUS proto-collaboration \cite{vahsen2020cygnus}, aims to employ a directional detector for rare event searches such as DM direct detection. Its concept consists in a Time Projection Chamber (TPC) filled with a He:CF$_4$ gas mixture at room temperature and atmospheric pressure with an amplification stage based on Gas Electron Multiplier (GEM) \cite{sauli2016gem} technology. GEM detectors are transparent to the light generated in the electrons' avalanches. Thanks to the scintillating properties of the gas \cite{margato2013scintillation,morozov2012scintillation} and to the amplification stage the readout is optical, comprised of scientific CMOS (sCMOS) or quantitative CMOS (qCMOS) cameras and photomultiplier tubes (PMTs). The high granularity of the amplification stage and readout enables precise three-dimensional reconstruction of particle tracks down to few mm while achieving low energy threshold below 1 keV$_{ee}$ (electron equivalent). \\
The collaboration is focusing on the development and construction of the demonstrator detector, CYGNO-04. It will consist in a TPC in back-to-back configuration of 0.4 m$^3$ of sensitive area with 50x80 cm$^2$ amplification area. The readout foresees 3 qCMOS cameras and 8 PMTs per side. The design foresees a concentrical structure with external shields of water and copper which contain an internal polymethyl methacrylate (PMMA) gas tight vessel which holds in the sensitive volume. It will be installed in the Hall F at underground laboratories of Gran Sasso (LNGS). The main goals are to prove the scalability of the detector technology and to assess the capability of control and prediction of the background which can provide reliable information on the physics reach of even larger detectors. Due to the latter, the realization of the internal structure of the detector with the least amount of material and with the highest radiopurity is of paramount importance to reduce the internal background contribution. At the same time, the correct electrical and mechanical behavior must be preserved in order to maintain a steady and trustworthy data taking. In this paper, different structures for field cage (FC) and cathode are described, tested and validated.

\section{Experimental Setup}
\label{sec:experimental_setup}
\begin{figure}[t]
    \centering
    \includegraphics[width=0.45\textwidth]{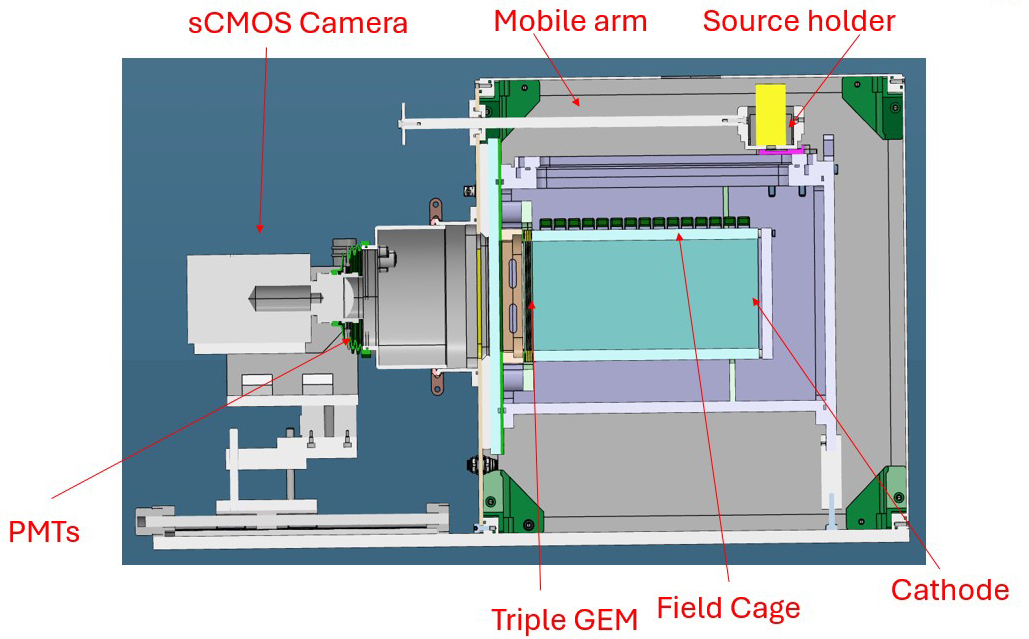}
    \includegraphics[width=0.45\textwidth]{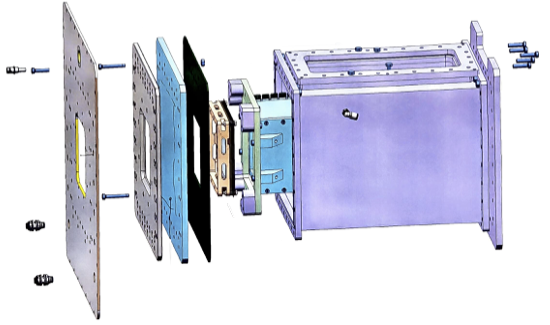}
    \caption{On the left is shown a cross section of the GIN detector design with the main components highlighted, on the right and exploded view of the internal part of the detector. }
    \label{img_gin_scheme}
\end{figure}

The detector utilized in the following tests is the Gaseous Imaging Nuclear recoil detector (GIN) built and utilised at the Laboratori Nazionali di Frascati (LNF) of Instituto Nazionale di Fisica Nucleare (INFN) . Its concept is based on the typical CYGNO technology design: a TPC filled with He:CF$_4$ gas mixture in a 60:40 ratio at atmospheric pressure and room temperature. It features an amplification stage based on a stack of three 50 $\mu$m GEM and the readout is optical. A sketch of the detector and an isometric view of its internal exploded structure are displayed in Figure \ref{img_gin_scheme}. The prototype main body is enclosed in a 3 mm thick aluminum Faraday cage with dimensions of 42 × 40 × 44 cm$^3$. Inside the Faraday cage, a gas tight PMMA vessel is positioned centered within with dimensions of 22 × 22 × 30 cm$^3$, corresponding to a volume of 14.52 liters. On top of the vessel, a 125 $\mu$m thick and 2 cm wide ethylene tetrafluoroethylene (EFTE) film serves as a window with high X-rays transparency down to few keV. A mobile source holder is placed above and can house radioactive sources and collimators used for calibration. A movable arm with handle outside the Faraday cage adjusts the position of the source in the longitudinal dimension of the sensitive area of the TPC. On the side of the readout, the PMMA vessel is sealed to a custom printed circuit board (PCB) through screws and O-ring. The PCB is designed to deliver the voltage input to the amplification stage and FC which are located inside the PMMA vessel and to mechanically support the amplification stage. The latter comprises three standard GEMs (50 $\mu$m thickness, 70 $\mu$m hole size and 140 $\mu$m pitch) spaced 2 mm each. The GEMs are numbered 1 to 3 with number 3 being the one furthest from the sensitive volume of the detector. The GEMs, the FC and the cathode inside the PMMA vessel determine the TPC sensitive volume where the readout area has dimensions of 10 × 10 cm$^2$, whilst the longitudinal drift length is of 23 cm.\\
A set of power supplies powers the electrical components. An ISEG HPn 500 705\footnote{\tiny\url{https://iseg-hv.com/files/media/Manual_HPS_350W_eng.pdf}}, with a maximum voltage of -50 kV and a maximum current of 7 mA powers the cathode electrode. A CAEN A1515TG\footnote{\tiny\url{https://www.caen.it/products/a1515b/}} High Voltage (HV) with floating channels with 1 kV supply maximum each is employed to provide voltage to the GEM stack and to the electrode of the FC closest to them.\\
The PMMA vessel is continuously flushed at a rate of 150 cc/min with a  mixture, obtained from bottles of pure gases in a ratio of 60:40 of He:CF$_4$. No recirculation of the gas is foreseen and the output gas is sent to an exhaust line connected to the external environment via a water-filled bubbler ensuring an over-pressure of approximately 3 mbar, relative to the external atmospheric pressure. \\
Outside the sensitive region on the side of the GEMs a stretched 150 $\mu$m thin polyethylene terephthalate (PET) foil acts as an optical window leading to a light-tight bellow connected to the optical system. This comprises a Hamamatsu Orca Fusion\footnote{\tiny\url{https://www.hamamatsu.com/content/dam/hamamatsu-photonics/sites/documents/99_SALES_LIBRARY/sys/SCAS0136E_C14440-20UP.pdf}} sCMOS camera centered in the bellow aperture and two Hamamatsu R1894\footnote{\tiny\url{https://www.alldatasheet.co.uk/datasheet-pdf/pdf/62693/HAMAMATSU/R1894.html}} PMTs located on the sides of the camera. The latter were not used during the tests. The Fusion camera is characterized by 2304 x 2304 pixels each with an extremely low noise of 0.7 electron root mean square (RMS) and a dimension of 6.5 x 6.5 $\mu$m$^2$. It is equipped with a Schneider Xenon lens with an aperture of $N = 0.95$ and focal distance set to 37 mm. In this configuration the optical system images an area of 11.5 x 11.5 cm$^2$ focused on the GEM3 plane with an effective pixel size of 50 x 50 $\mu$m$^2$ and an optical acceptance of $\Omega = 9.2 \times 10^{-4}$.

\subsection{Tested components}
\label{sec:tested_components}
In this section, we describe the components that were tested with the GIN prototype.\\
In order to reduce the material required for the FC, the chosen design consists in a sheet of 45 $\mu$m thick of PET of 40.3 x 23 cm$^2$ provided by the SERIGROP company. Eleven strips of Cu 35 $\mu$m thick and 1 cm wide are deposited along the long side with a pitch of 2 cm. The PET in the sheet can be replaced by Kapton with no mechanical and electrical variation. Once the sheet is folded to close on itself it becomes a parallelepiped 23 cm long and with a square base of 10 cm of side, perfectly acting as a FC for the GIN detector. The 3 extra mm are used to clip the foil on each side to permit its correct stretching. While this provides a reduced amount of material, it is not mechanically self supporting. The supporting structures and two different cathodes tested with the GIN prototype are described in the following.

\begin{figure}[t]
    \centering
    \includegraphics[width=0.45\textwidth]{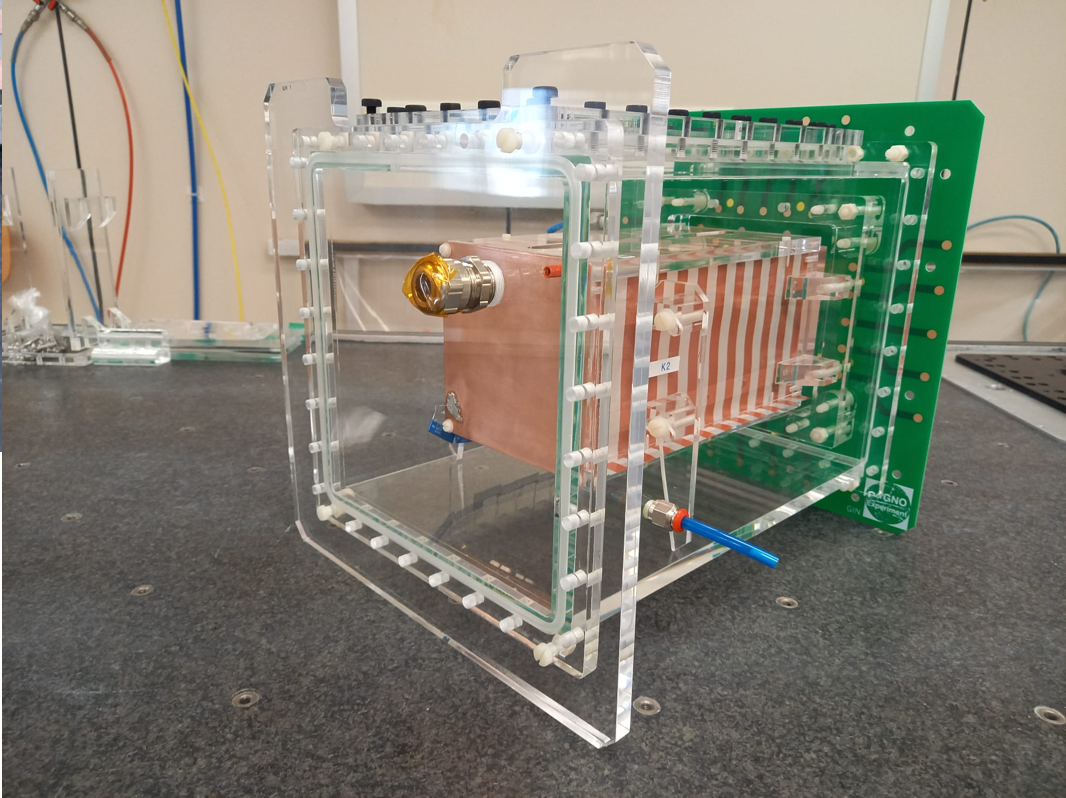}
    \includegraphics[width=0.253\columnwidth]{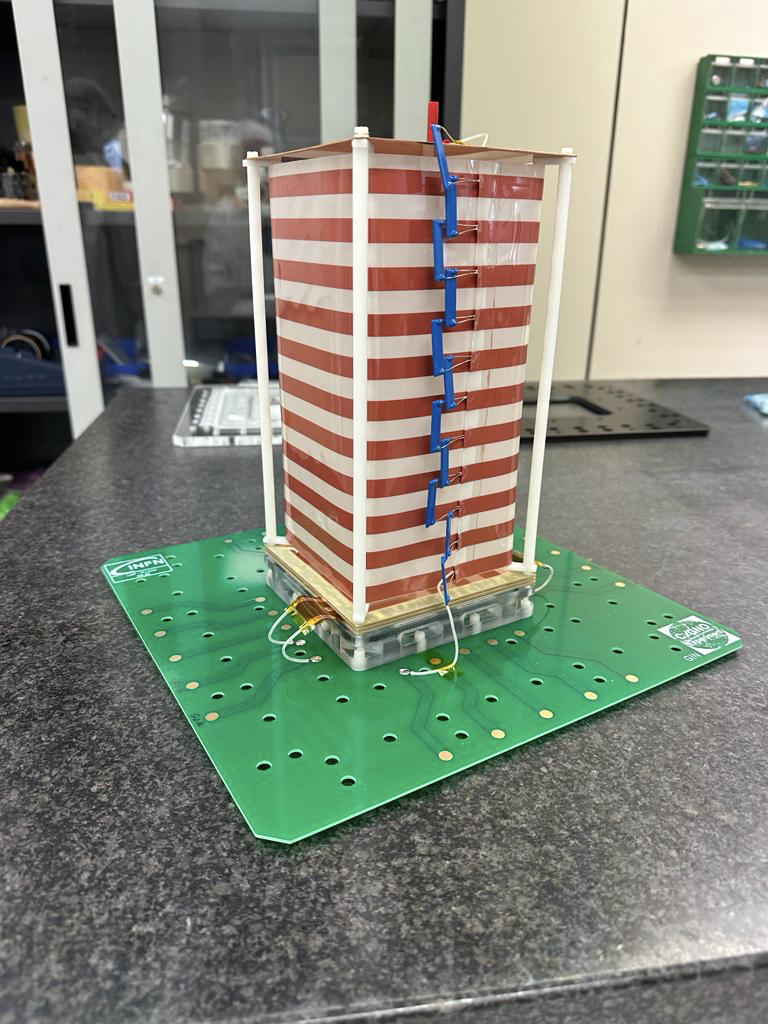}
    \caption{On the left, the F1 field cage structure encased in the PMMA gas tight vessel, while on the right the F2 is shown. Both are equipped with cathode C1.}
    \label{img_P0_pic}
\end{figure}

\begin{itemize}
    \item \textbf{Field cage 1 (F1)} — In F1 the FC foil is glued  to polyvinyl chloride (PVC) slabs using 3M adhesive tape. Four slabs are held together through PMMA bridges directly attached to the PMMA vessel. The resistors electrically connecting the FC rings are soldered to the rings on the clipped corner along the long side of the FC. A picture of this structure is shown on the left of Figure \ref{img_P0_pic}. 

    \item \textbf{Field cage 2 (F2)} — The F2 is made by rolling the FC sheet around four DELRIN pillars, one at each corner. The foil is glued to itself after one complete turn. Four additional DELRIN pillars connect the cathode to the GEM support structure. With respect to F1, no additional PMMA and a lower amount of glue are required while the shape of the FC sheets results slightly rounded on the corners. A picture of F2 is shown on the right in Figure \ref{img_P0_pic}.

    \item \textbf{Field cage 3 (F3)} — The F3 improves the design of F2 with larger and sturdier pillars and other elements of support which are enough to sustain the full FC sheet foreseen for the CYGNO-04 detector. While the pillar is thicker than F2, in the region inside the FC sheet the foils is stretched and kept attached to the pillar round-shaped screws in order to limit the material inside the sensitive region. In addition, the larger size of the external pillar is exploited to house the low radioactive surface-mount device (SMD) resistors \cite{pesudo2023smd}. The whole structure is made in Nylon6 which is very low radioactive \cite{arpesella2002borexino,zuzel2003nylon}. A picture of F3 is shown on the left in Figure \ref{img_P2_pic}.

    \item \textbf{Copper cathode (C1)} — C1 is a simple flat copper plate with dimensions of 10 × 10 cm$^2$ and a thickness of 1 mm. Copper can be procured extremely radiopure and requires extremely simple construction. Figure \ref{img_P0_pic} shows two different FC structures both with the C1 cathode installed.

    \item \textbf{Aluminized mylar cathode (C2)} — This cathode design is inspired by the one used by the DRIFT Collaboration~\cite{battat2015reducing}. It is made by a stretched foil of aluminized mylar with a thickness of 0.9 $\mu$m. Whilst of complex installation, this cathode is designed to provide a method for back-to-back TPC of tagging the radon-progeny-recoils, nuclear recoils induced by alpha decays of radioactive elements attached to the cathode, which can emulate DM signal. The extremely fragile cathode is powered by soldering the HV wire to a copper tape glued on the side of the cathode. A picture of C2 is shown on the right in Figure \ref{img_P2_pic}.

    \begin{figure}[t]
        \centering
        \includegraphics[width=0.35\textwidth]{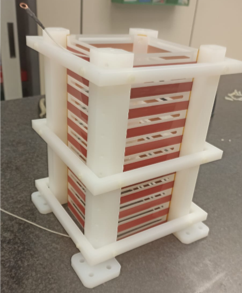}
        \includegraphics[width=0.318\textwidth]{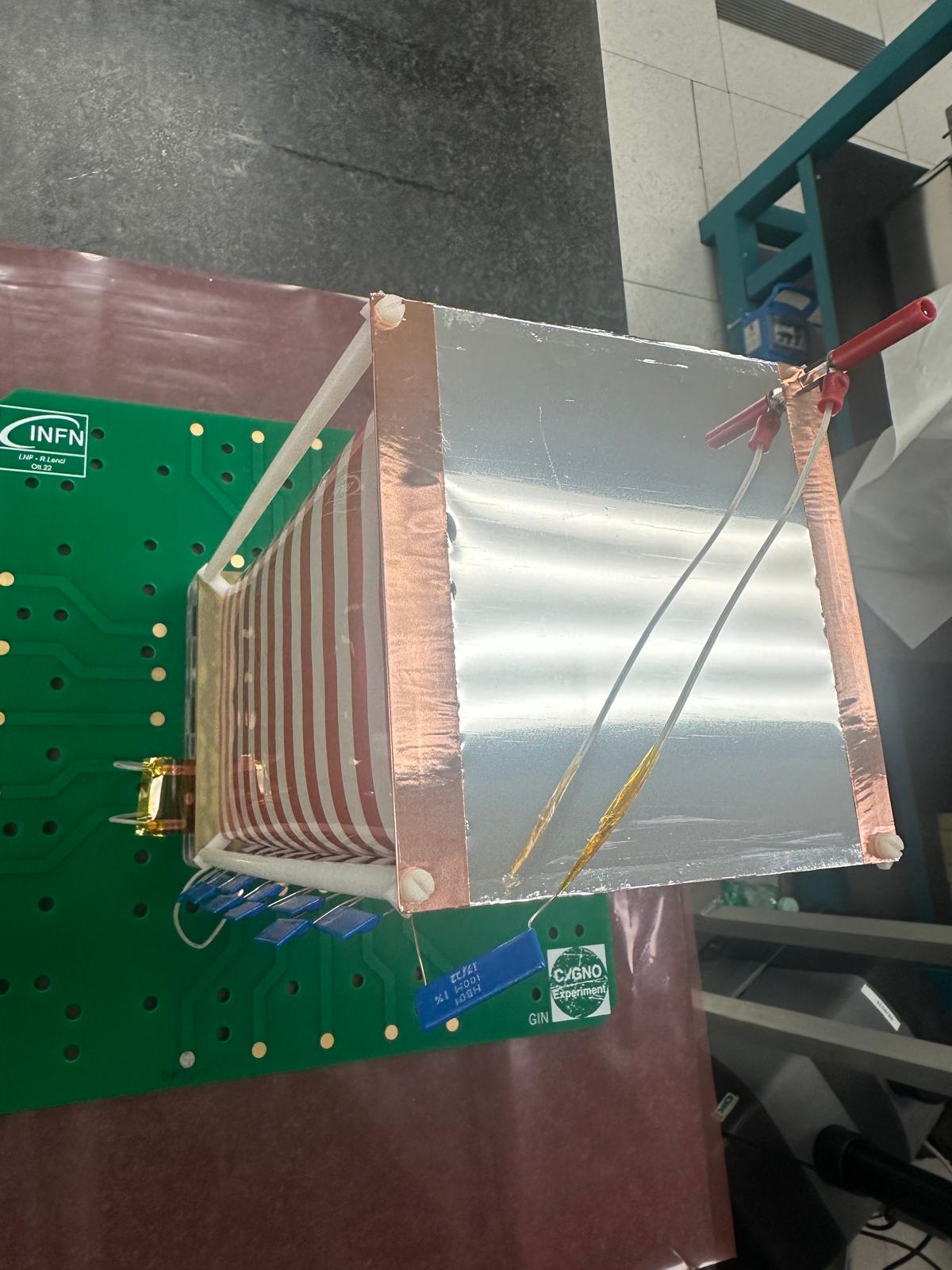}
        \caption{On the left the F3 field cage structure, while on the right the F2 equipped with cathode C2 is shown.}
        \label{img_P2_pic}
    \end{figure}
\end{itemize}
\begin{table}[t]
\centering
\begin{tabular}{ccc}
\toprule
Configuration name & Field cage structure & Cathode \\
\midrule
P0  & F1  &  C1 \\
P1  & F2  &  C1  \\
P2  & F2  &  C2  \\
P3  & F3  &  C1 \\
\bottomrule
\end{tabular}
\caption{Different configuration of FC structure and cathode tested in this work. Names of FC and cathodes refer to the description in the text.}
\label{table_P}
\end{table}
The components just described were tested in different configurations summarized in Table \ref{table_P}.

\section{Experimental Method}
In this section, we provide a detailed description of the experimental method used to test the components. We established four separate measurements to ensure stability and electrical properties of the FC and cathode assembly, based on the uniformity of the drift field. The voltage setting employed for all configurations and data sets corresponds to the one estimated for the CYGNO-04 detector: 1 kV/cm of drift field, voltage across each GEM of 440 V and a transfer field between GEMs of 2.5 kV/cm. In all the measurements data were acquired with the standard acquisition DAQ model. In particular, a copy of PMT signals is passed to a discriminator to generate a digital outputs. The camera hardware provides a digital signal which is positive whenever more than 1 sensor row is active. When a coincidence of the camera signal with both PMT ones occurs, a trigger is fired which allows to save the camera image and all waveforms above discriminator threshold for until the camera signal is on. The analysis is performed solely on the camera images for the purposes of this study.

The images acquired are passed to the standard reconstruction code of the CYGNO experiment, which selects the pixels corresponding of each signal (soft electrons, muons, alpha tracks) and estimates variables like energy deposited, size of the cluster and more. A more detailed description of the algorithm operation can be found in \cite{dho2024ely,pains2023idbscan,dimarco2020identification}.
\subsection{Stability}
To ascertain that the structure tested was stable in a relatively long period, the detector assembly was kept on for a month. The gas was monitored with pressure, temperature and humidity sensor. After assembly, the detector was flushed until the humidity value dropped below 2\% RH, which is considered low enough to have negligible effect on the light yield of the detector. The spark rate limit was set to $1.5 \times 10^{-3}$ Hz to ensure the data quality. In this configuration, any data for analysis can be acquired. During the month-long period, the currents readout by the HV system were monitored in order to identify the spark rate. In addition, thanks to the different current channels and the sCMOS camera images, in case of sparks or defects it was possible to determine the region or piece culprit of the instability. These kind of faulty behavior was also tested by forcing the drift field up to 1.5 kV/cm, well above the expected operating conditions, to verify the solidity of the structure

\subsection{Collection efficiency}
\begin{figure}[t]
    \centering
    \includegraphics[width=0.50\columnwidth]{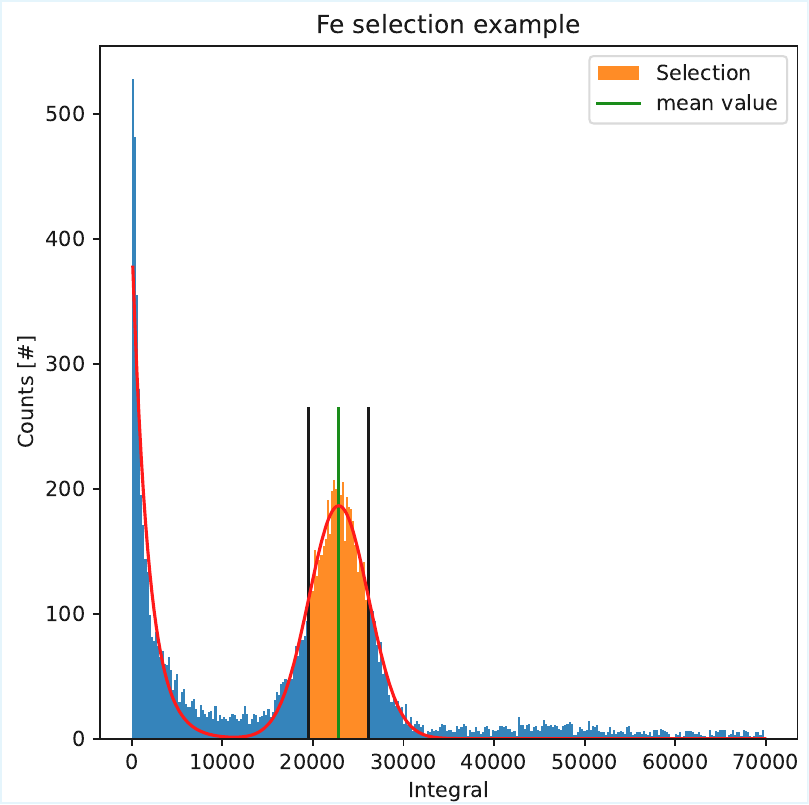}
    \caption{Example of Fe clusters selection with the GIN detector. In orange, the selected iron clusters used for evaluating the collection efficiency.}
    \label{img_sc_integral_calibration}
\end{figure}

In order to test the regularity of the drift field in the drift direction (z coordinate henceforth) we employed a 480 kBq $^{55}$Fe source positioned on the source holder to exploit the 5.9 keV X-ray emission converted in the sensitive region. As the distance traveled by the 5.9 keV photoelectron in the gas is $\mathcal{O}$(100) $\mu$m and the diffusion is a slightly larger, the x-y projection of this signal as seen by the sCMOS camera appears round. Data with the $^{55}$Fe source is 
acquired with an exposure of 150 ms to ensure enough signal is present in each single picture without risking large pileup.
The acquired data is analyzed by the reconstruction algorithm, the signals identified and a spectrum of the \textit{Integral} of the clusters is obtained. The Integral of a cluster is the sum of the intensity of the pixels belonging to it, representing a quantity proportional to the energy deposited in the gas. Geometrical and quality cuts are applied to remove noise and unsuitable events. The geometrical cut excludes events outside the FC area. The quality cut selects clusters with a slimness (the ratio between their minor and major axes) greater than 0.6, since tracks from the source are expected to be round. An example of an integral spectrum is shown in Figure \ref{img_sc_integral_calibration}.
The Integral distribution of the selected clusters is fitted with an exponential function for noise and a Gaussian for the signal. The mean of the Gaussian corresponds to the 5.9 keV $^{55}$Fe peak. For each run, clusters with an Integral within $\pm 1\sigma$ from the mean  of the $^{55}$Fe Gaussian are selected to count the number of detected signals. Such a choice guarantees that the selection range is adapted dynamically to follow the variation of the $^{55}$Fe population's distribution. Moreover, it ensures the selection of a region with high signal to noise ratio. The collection efficiency is studied as the number of signals counted as a function of the distance in drift distance.
\subsection{Diffusion}
To further test the behavior of the drift field as a function of the z coordinate, the diffusion properties in the gas are measured. In fact, diffusion processes can be modeled \cite{blum2008particle} as:
\begin{equation}
    \sigma^{2} = \sigma^{2}_{0} + \xi^2 \cdot z
\end{equation}

The first term, $\sigma_0^2$, accounts for GEM intrinsic diffusion including possible saturation deformation terms, the original track dispersion and any term independent of z. The $\xi$ term describes drift-induced diffusion along $z$. The value of $\sigma^2$ is measured from the round shape of the $^{55}$Fe signals by calculating the RMS of the distribution of pixel intensity in the minor axis of the cluster. Thus, $\xi$ can be evaluated from the $^{55}$Fe data. Given the larger importance of a more precise z determination of the position of the $^{55}$Fe signal, we employed a collimator with a thin slit perpendicular to the drift direction. This reduced the uncertainty on the z down to 0.5 cm \cite{pinci2025gem}. Garfield simulation of the diffusion coefficient of our gas mixture returns a value of  $\xi_{\text{sim}} = (110 \pm 1)\ \frac{\mu\text{m}}{\sqrt{\text{cm}}}.$
which will be compared with experimental data.
The term $\sigma_0^2$ will instead be ignored as its dependence is related to the GEM diffusion which is outside the scope of this paper.

\subsection{Disuniformity in x-y}
The uniformity of the electric field in the x-y plane, the plane parallel to the amplification stage is of paramount importance to ensure all the sensitive volume collects charge equally. In previous CYGNO prototypes \cite{LIMEpaper}, a rather strong disuniformity between the top and the bottom part of the image was found of the order of 30\%. Therefore, the CYGNO-04 requirements allow a maximum top-bottom (T-B) and left-right (L-R) disuniformities of 10\%.

The measurement consists in the acquisition of images with the sCMOS camera of the natural radioactivity of the laboratory. Exploiting soft electrons and muons which can randomly cross the detector in any position, a x-y hitmap can be created. In particular, the images are passed to the reconstruction algorithm which selects the pixels belonging to all events detected. Each pixel of the hitmap is filled with the sum of the content of all pixels with that coordinate which belonged to a signal cluster divided by the times that pixel was part of a cluster. This way, the intensity of the hitmap is normalized in intensity to the average light collected, possibly highlighting disuniformities. Images where the total amount of light exceeds a threshold of 10$^6$ times the average light collected in a image are rejected as they may contain sparks. The hot spots which occur always in the same position are removed as well. Every other kind of event is accepted and considered in the analysis.

\begin{figure}[t]
    \centering
    \includegraphics[width=0.55\columnwidth]{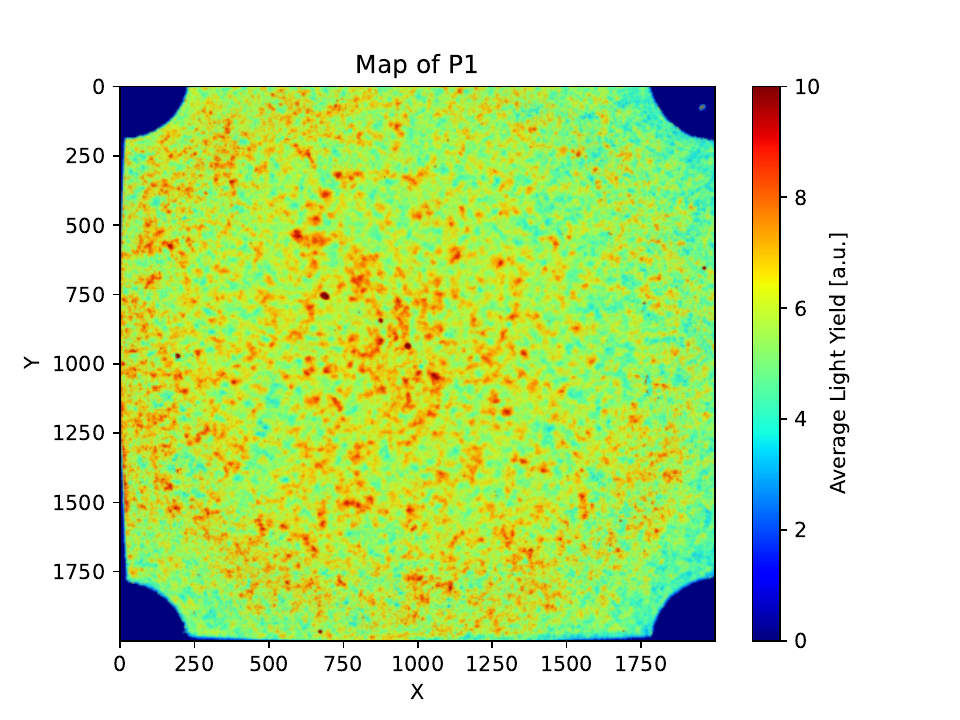}
    \caption{FC map of the P1 configuration.}
    \label{img_map}
\end{figure}
Figure \ref{img_map} shows an example of a hitmap of P1. The original collected images have dimension of 2304 x 2304 pixels, comprising regions outside the GEM active area. Those regions are cut away and the final image dimension is then reduced to 2000 x 2000 pixels to perfectly represent the 10 x 10 cm$^2$ area of the readout and FC structure.
Besides qualitative estimation of dead spots, a quantitative analysis is performed by rebinning the image to a 10 x 10 pixel square. The intensity of the five outermost column of pixels on the left and right side area (corner excluded) is averaged and compared to each other to estimate L-R disuniformity.

In order to evaluate possible disuniformities which are T-B or L-R symmetric, we performed an analysis based on Bessel's functions \cite{arfken2011mathematical} decomposition, which is the classical way to describe the vibrational modes in the circle. First of all, a Gaussian filter with a sigma of 2 pixels is applied to remove single-pixel fluctuations. The square map is then reduced to a circle with a radius of 2000 pixels, excluding the corners, and it is normalized to its average. The border of the circle, with a width of 1 pixel, is set to 0 to enforce the Dirichlet's boundary conditions. A least squares fit with the real basis of the Bessel's function is performed on the map to determine the coefficients of the expansion. For simplicity, we limited the decomposition to $m_{max} = 3$ and $n_{max}= 4$, as we are interested in large scale disuniformities which could be induced by the FC rather than single pixel differences. The sum  of the modulus squared of the decomposition coefficients $\sum_n |a_{nm}|^2$ for the angular order $m$ represent the disuniformity on that given order. Based on the previous analysis, we consider acceptable configurations which show disuniformities under 10\% for every angular order. The results are compared with a reference map, where a 10\% disuniformity between the minimum and the maximum for each angular order is enforced adding to the identity map the first mode of the order multiplied for a proper constant. 

\section{Results}
For all the configurations listed in Table \ref{table_P}, the tests and measurements discussed in the previous Section were undertaken.\\
\begin{figure}[t]
    \centering
    \includegraphics[width=0.45\columnwidth]{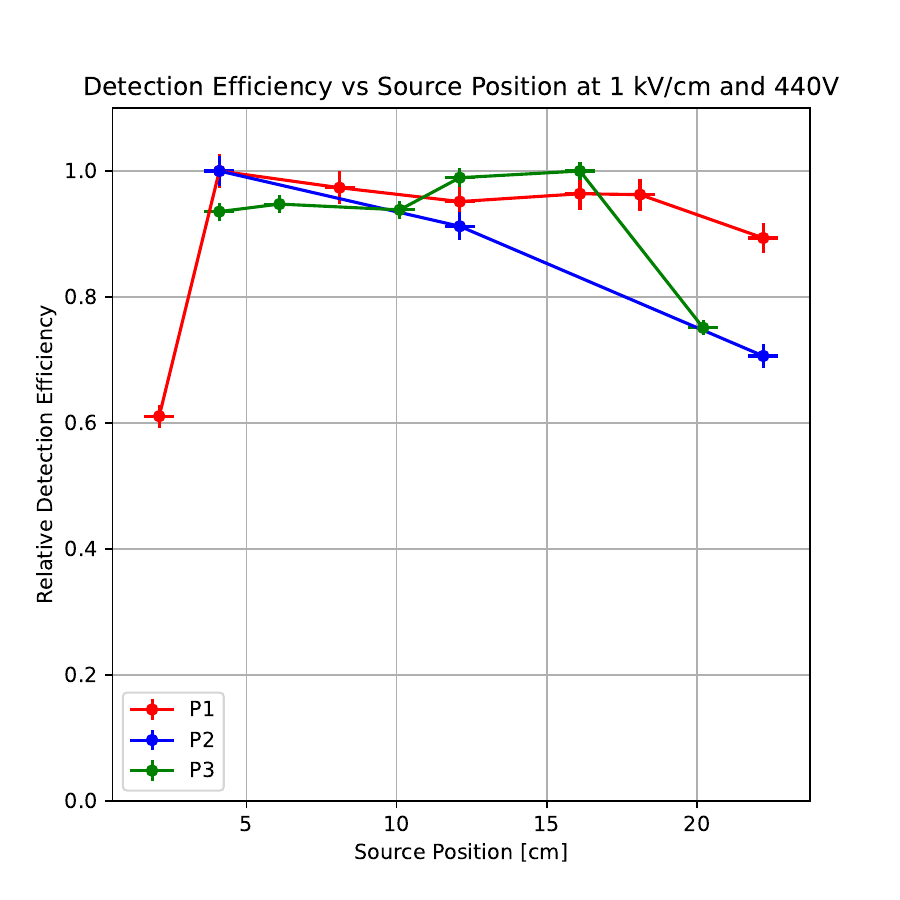}
    \includegraphics[width=0.45\columnwidth]{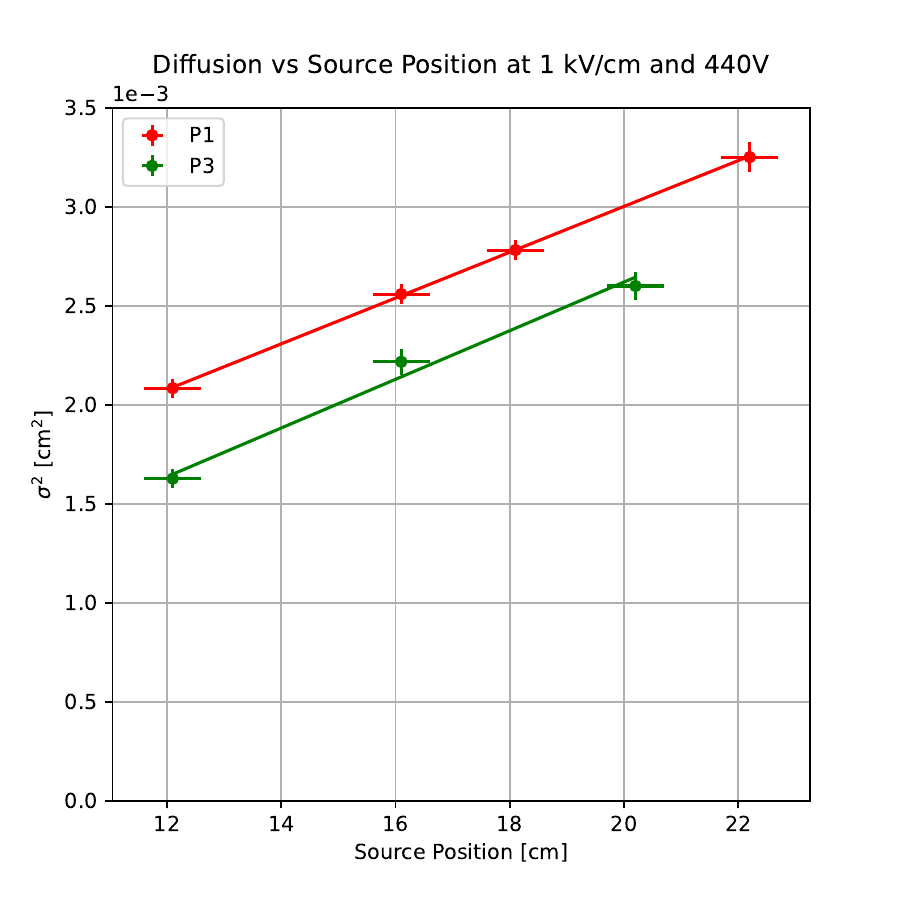}
    \caption{(left) Relative detection efficiency as a function of the source position. (right) Diffusion as a function of the source position. The points represent the measured values. The colored lines in the right plot show the best-fit curves.}
    \label{img_results_short}
\end{figure}
In terms of stability, the configurations P1 and P3 had no issues and passed all the tests. For what concerns P0, after few days of operation, it started discharging on the sides of the FC sheet and along the PVC structure. We tried to reduce the sharpness of the sheet without improving the operative conditions. In fact, within a couple of days the maximum drift field achievable had to be decreased. We suppose that the glue and the surface of the PVC can foster the creation of surface currents, which can induce sparks.\\
Regarding P2, the C2 cathode posed no issue in the first weeks. However, in time the maximum voltage reachable dropped down to a point where no data could be taken. For this reason, the diffusion test was not performed on P2. Further inspection revealed that the connection between the copper tape and the cathode sheet was degraded. This suggests that an improved method to perform the electrical connection is required to employ this cathode.\\

On the left panel of Figure \ref{img_results_short}, a comparison of the relative detection efficiencies for three configurations is displayed. The efficiency is normalized to the maximum recorded value for each configuration to obtain the relative value presented in the plot. For configuration P1 and P3, the efficiency is pretty constant along the z dimension and above 90\% of the relative value. The exception is represented by significant drops at the border of the detector. These are caused by the angular dispersion of the emitted radiation which is produced in a cone that partially exits the active region of the detector when the source is positioned too close to the GEMs or the cathode. On the other hand, P2 demonstrates a decreasing efficiency with larger z.
It might be an indication of the field not be constant along the drift direction. Despite the humidity measurement being a quite reliable reference for the gas quality, we cannot be sure that other electronegative impurities did not affect our gas mixture during the specific data taking. \\

The right panel of Figure \ref{img_results_short} shows the measured diffusion values $\sigma^2$ on the y-axis as a function of the source position on the x-axis. Data only with z larger than 10 cm was utilized as the constant term and possibly space charge effect on the GEM holes can modify the shape of the $^{55}$Fe signal \cite{pinci2025gem}. A linear fit performed on the data to extract the diffusion parameter $\xi$. The result of the P1 and P3 configuration are respectively $
\xi_{\text{P1}} = (108 \pm 3) \frac{\mu\text{m}}{\sqrt{\text{cm}}}
$
and 
$\xi_{\text{P3}} = (111 \pm 5) \frac{\mu\text{m}}{\sqrt{\text{cm}}}
$.
The two values are compatible with each other and with the expected value attained with the Garfield simulation, confirming the correct behavior of the field.\\

\begin{figure}[t]   
    \centering
    \includegraphics[width=0.45\columnwidth]{img/results/P1_cut.pdf}
    \includegraphics[width=0.45\columnwidth]{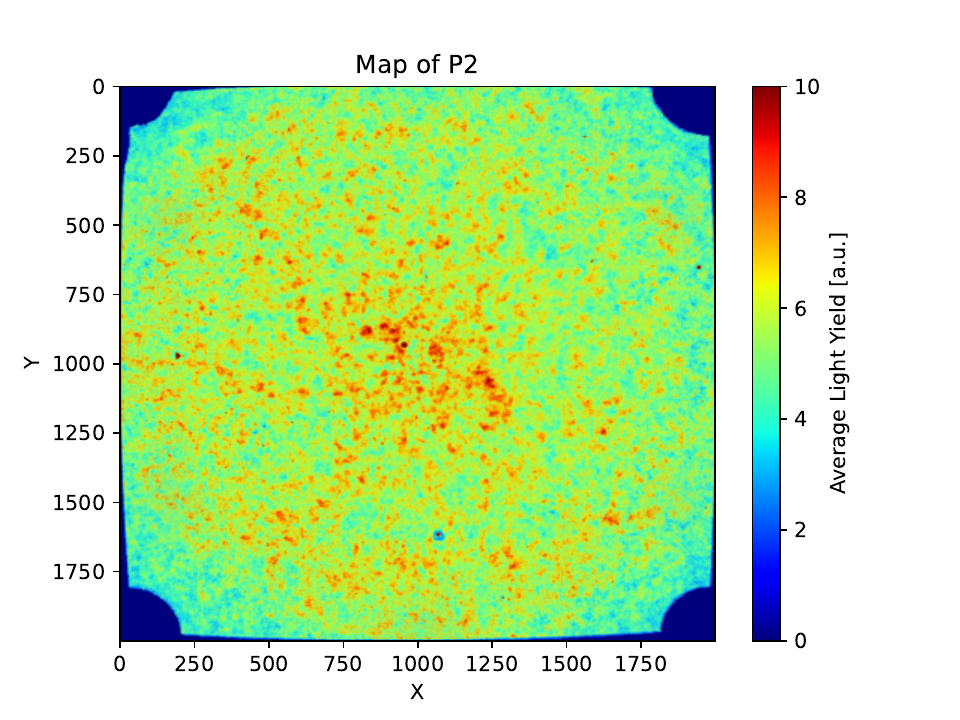}
    \includegraphics[width=0.45\columnwidth]{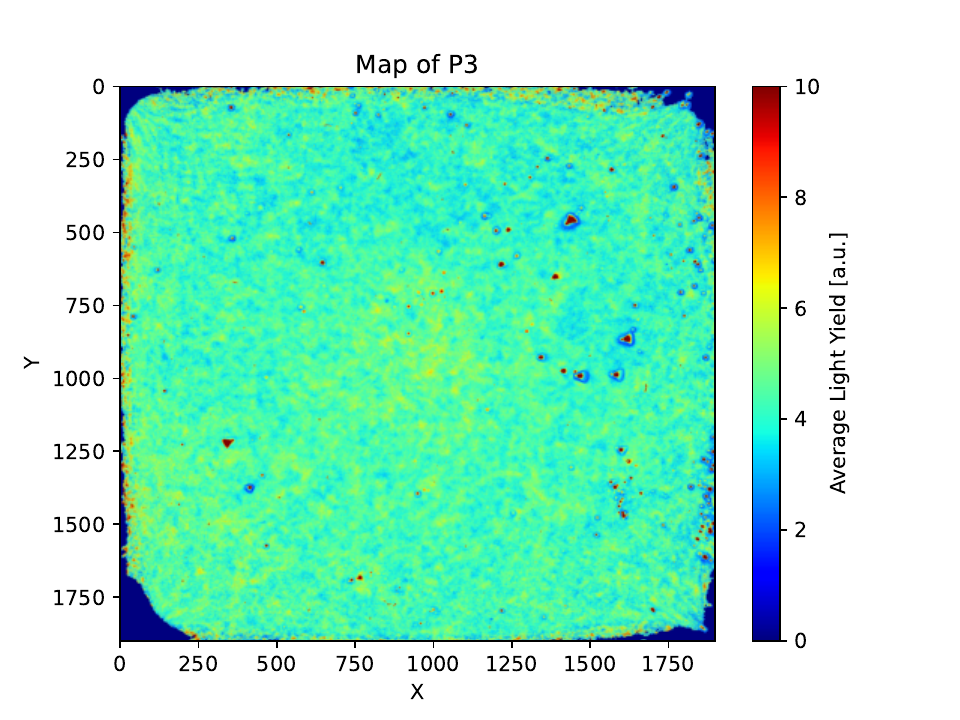}
    \caption{Field Cage Maps of the P1, P2, and P3 configurations.}
    \label{img_maps}
\end{figure}

Figure \ref{img_maps} shows the hitmaps of the three configurations that passed the conditioning phase. P1 and P2 share the same FC structure F2 and indeed have similar blind spots on the four corners of the sensitive region. These are caused by the internal DELRIN pillar which block completely the electron transmission. P1 exhibits disuniformity of 0.46 $\pm$ 0.04\% T-B towards the bottom and 6.25 $\pm$ 0.05\% L-R towards the left. On the other hand, P2 has 0.37 $\pm$ 0.04\% T-B and a L–R of 2.71 $\pm$ 0.04\%, with the top and left regions favored, respectively.

Concerning the configuration P3, the hitmap reveals that, thanks to the use of internal screws instead of pillars, the dead area of the active region is smaller. Close to the corners, it is possible to notice that the hitmap, and therefore the reconstructed signal clusters, are deformed with what appears to be a drag towards the tip of the corners. This is due to the distortion of the drift electric field closing on the GEM corners which are slightly more outwards than the corner of the field cage. Based on the pixel dimension, the distortion region occupies about 6\% of the total area of GIN. However, since the F3 structure is design to be already capable of supporting the FC sheet of the dimension required for CYGNO-04, the area affected by distortion in a  readout with size of 50x80 cm$^2$ is about 0.1\% and thus considered negligible. Concerning the disuniformity P3 presents a T-B of 1.89 $\pm$ 0.05\% and a L–R of 1.17 $\pm$ 0.05\%, with the lower and left regions favored. Results are summarized in Tab. \ref{table_results}.
All values are well below the 10\% threshold, and therefore meet the requirements for CYGNO-04.

\begin{table}[t]
\centering
\small
\begin{tabular}{ccccc}
\toprule
Configuration  & Stability Test & Diffusion Parameter [$\frac{\mu m}{\sqrt{cm}}$] & T-B Disuniformity & L-R Disuniformity  \\
\midrule
P0  & Not Passed  &  &  & \\
P1  & Passed  &  108 $\pm$ 3 & 0.46 $\pm$ 0.04 \% (B) & 6.25 $\pm$ 0.05 \% (L)\\
P2  & Passed  &   & 0.37 $\pm$ 0.04 \% (T) & 2.71 $\pm$ 0.04 \% (L) \\
P3  & Passed  &  111 $\pm$ 5 & 1.89 $\pm$ 0.05 \% (B) & 1.17 $\pm$ 0.05 \% (L)\\
\bottomrule
\end{tabular}
    \caption{Summary of the results. The letters in round brackets indicate the favored side.}
\label{table_results}
\end{table}

\begin{figure}[t]
    \centering
    \includegraphics[width=0.60\columnwidth]{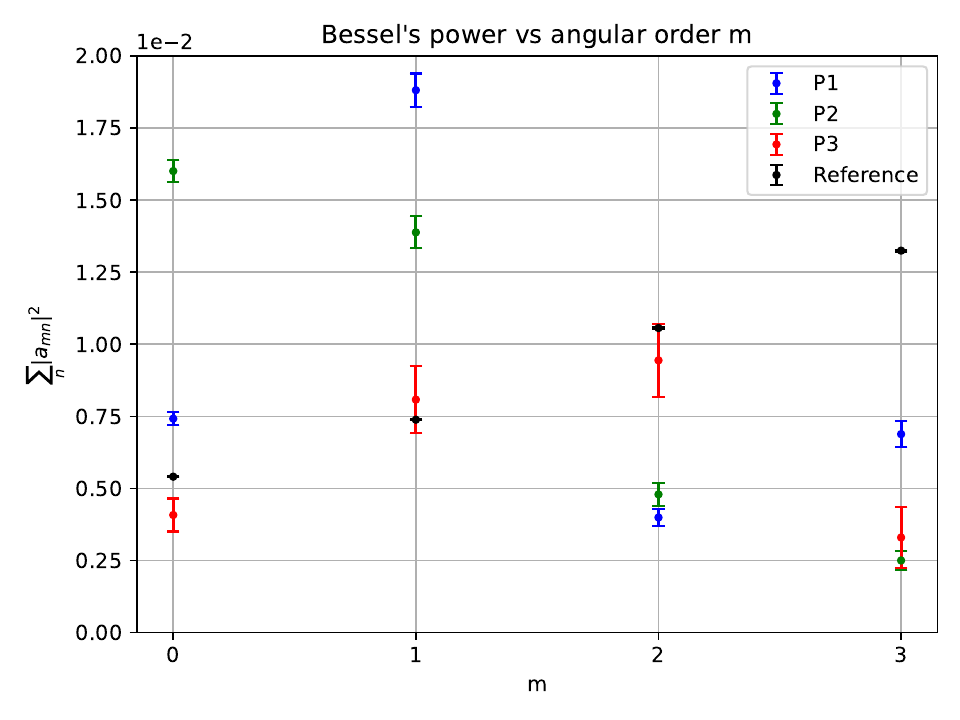}
    \caption{Power spectra of the decomposition in Bessel's functions.}
    \label{img_besssel}
\end{figure}

Figure \ref{img_besssel} shows the fitted power spectra of the decomposition is Bessel's functions for the three configurations and for the reference map. P1 and P2 show rather high radial and dipolar disuniformities, exceeding the acceptable threshold for those modes. Radial disuniformities could be partially due to natural optical vignetting, where the light intensity decreases radially from the optical centre of the system. However, we expect vignetting to give equal contribution to all the configurations. P3 shows slightly higher quadrupolar disuniformity than the other two configuration, although compatible within 1 $\sigma$ with the reference map. For higher order disuniformities, all the configurations are significantly below the threshold. The P3 configuration is the only one which satisfies the further constraints applied from requirements. Summary of the results of Bessel's decomposition analysis can be found in Tab. \ref{table_bessel}.

\begin{table}[t]
\centering
\small
\begin{tabular}{ccccc}
\toprule
Configuration  & $m = 0$ & $m = 1$ & $m = 2$  & $m = 3$   \\
\midrule
P1  & $ 7.4 \pm 0.2 \times 10^{-3}$  & $ 1.88 \pm 0.06 \times 10^{-2}$ & $ 4.0 \pm 0.3 \times 10^{-3}$ & $ 6.9 \pm 0.5 \times 10^{-3}$\\
P2  & $ 1.60 \pm 0.04 \times 10^{-2}$  & $ 1.39 \pm 0.06 \times 10^{-2}$  & $ 4.8 \pm 0.4\times 10^{-3}$ & $ 2.5 \pm 0.3 \times 10^{-3}$ \\
P3  & $4.1 \pm 0.6 \times 10^{-3}$  &  $ 8 \pm 1 \times 10^{-3}$ & $ 9.5 \pm 1 \times 10^{-3}$ & $ 3 \pm 1 \times 10^{-3}$\\
Reference  & $5.41 \pm 0.02 \times 10^{-3}$  &  $ 7.38 \pm 0.02 \times 10^{-3}$ & $ 1.056 \pm 0.003 \times 10^{-2}$ & $ 1.325 \pm 0.003 \times 10^{-2}$\\
\bottomrule
\end{tabular}
    \caption{Summary of Bessel's decomposition analysis.}
\label{table_bessel}
\end{table}

\section{Conclusions}
In the context of the directional direct search for WIMP-like DM signature, the CYGNO experiment is moving towards the construction of the demonstrator detector. This requires the employment of radiopure materials and minimal structures for the internal elements of the detector in order to reduce background. At the LNF laboratories with the use of the GIN prototype a FC based on a thin PET sheet with strips of copper deposition was tested with different supporting structures and cathode technologies. In order to verify the mechanical and electrical properties, each assembly was tested evaluating the stability performances along a month of operation. The correct response and drift electric field were tested along the drift direction by measuring the gas diffusion and the collection efficiency employing an $^{55}$Fe source. The field uniformity in the plane parallel to the amplification stage was tested by generating an intensity map of the response to natural radioactivity in order to spot disuniformities. 
The structure where the FC sheet is glued to PVC supports failed to pass stability tests due to discharges along the FC. On the contrary, the other FC structures demonstrated excellent performances in terms of stability and uniformity of the response along the drift direction and orthogonally. The aluminized mylar cathode worked correctly for a couple of weeks until the electrical connection got damaged. The uniformity map showed promising results, but the powering technique needs to be improved for it to be installed in a larger and long term detector.\\
In particular, the configuration P3 has the lowest dead area in the sensitive region due to construction material and it is already designed so that it can withstand a FC foil of the dimension necessary for the CYGNO-04 detector. Thus, the P3 configuration is validated and considered optimal for the realisation of CYGNO-04.

\acknowledgments
 This project has received fundings under the European Union’s Horizon 2020 research and innovation programme from the European Research Council (ERC) grant agreement No 818744. We want to thank General Services and Mechanical Workshops of Laboratori Nazionali di Frascati (LNF) and Laboratori Nazionali del Gran Sasso (LNGS) for their precious work for technical support. This work is (partially) supported by ICSC – Centro Nazionale di Ricerca in High Performance Computing, Big Data and Quantum Computing, funded by European Union – NextGenerationEU.


 \bibliographystyle{JHEP}
 \bibliography{biblio}
\end{document}